\documentclass[twocolumn,aps,prl,floatfix,superscriptaddress,longbibliography]{revtex4}
\pdfoutput=1 
\usepackage{amsmath,amssymb,eucal,graphicx,float,epstopdf}
\usepackage{subfigure}
\usepackage{comment}
\usepackage[utf8]{inputenc}
\usepackage[colorlinks=true, urlcolor=blue, anchorcolor=blue, citecolor=blue,filecolor=blue,linkcolor=blue,menucolor=blue]{hyperref}

\begin{document}
\title{Intermittency in Wind-Driven Fires}

\author{Laurent Hébert-Dufresne}
\affiliation{Vermont Complex Systems Institute and Department of Computer Science, University of Vermont, Burlington, VT 05408, USA}
\affiliation{Complexity Science Hub, 1030 Vienna, Austria}
\affiliation{Santa Fe Institute, 1399 Hyde Park Road, Santa Fe, NM 87501, USA}
\author{Aanjaneya Kumar}
\affiliation{Santa Fe Institute, 1399 Hyde Park Road, Santa Fe, NM 87501, USA}
\author{S. Redner}
\affiliation{Santa Fe Institute, 1399 Hyde Park Road, Santa Fe, NM 87501, USA}

\begin{abstract}
  We construct a wind-driven forest-fire model in one dimension in which a fire can jump gaps between trees to ignite disjoint downwind forests. The size of a gap that a fire can jump depends on the fire intensity, which increases as the fire propagates through trees and diminishes as the fire jumps gaps.  Trees grow on empty sites at rate $r$ and lightning strikes each site with rate $f$.  When $f\ll r/L$, where $L$ is the system length, lightning is sufficiently rare that quasi-deterministic dynamics arises where all trees are consumed when a lightning-induced fire occurs.  For $f\gg L^{-\mu}$ with $\mu\approx 0.8$, lightning is sufficiently frequent that a steady state is reached, but with unexpected behaviors for the fire- and gap-size distributions. Intermittency arises in between these regimes, with coexisting temporal domains of deterministic and chaotic dynamics.
\end{abstract}

\maketitle

\paragraph*{\textbf{Introduction:}} Forest-fire models based on self-organized criticality provide minimalist and rich descriptions for wildfire propagation~\cite{Bak1989ForestFire,GrassbergerKantz1991,DrosselSchwabl1992,Drossel1993exact,ClarDrosselSchwabl1996,SchenkDrosselSchwabl1999,SchenkDrosselSchwabl2001,Pruessner2002,vanDenBergJarai2005,hebert2018edge,Palmieri2020}. In these models, each of $L$ lattice sites can be empty, occupied by a tree, or occupied by a burning tree.  In the version of the model most relevant to this work, the dynamics consists of the following steps: At rate $r$, empty sites turn into trees. At rate $f$, lightning strikes a random site.  If the site is empty, nothing happens. If the site contains a tree, it and the  contiguous forest instantaneously burn and become empty. These appealingly simple rules lead to stationarity, in which forests are sporadically consumed by lightning-induced fires and continuously replenished by tree growth. When $f\ll r$, the probability $P(F)$ for a fire of size $F$ typically has a power-law tail, $P(F)\sim F^{-\nu}$, with the exponent $\nu$ between $1$ and $2$, depending on model details. 

However, real wildfires do not spread exclusively through neighboring trees.  A sufficiently intense fire may jump gaps between trees where little combustible material is present.  This intergap propagation is often facilitated by wind, which arises either externally or intrinsically from air currents generated by the fire itself.  Motivated by this observation, we formulate a model for a fire to jump gaps and spread to trees in disjoint forests.

\begin{figure}[ht]
\centering
\includegraphics[width=0.4\textwidth]{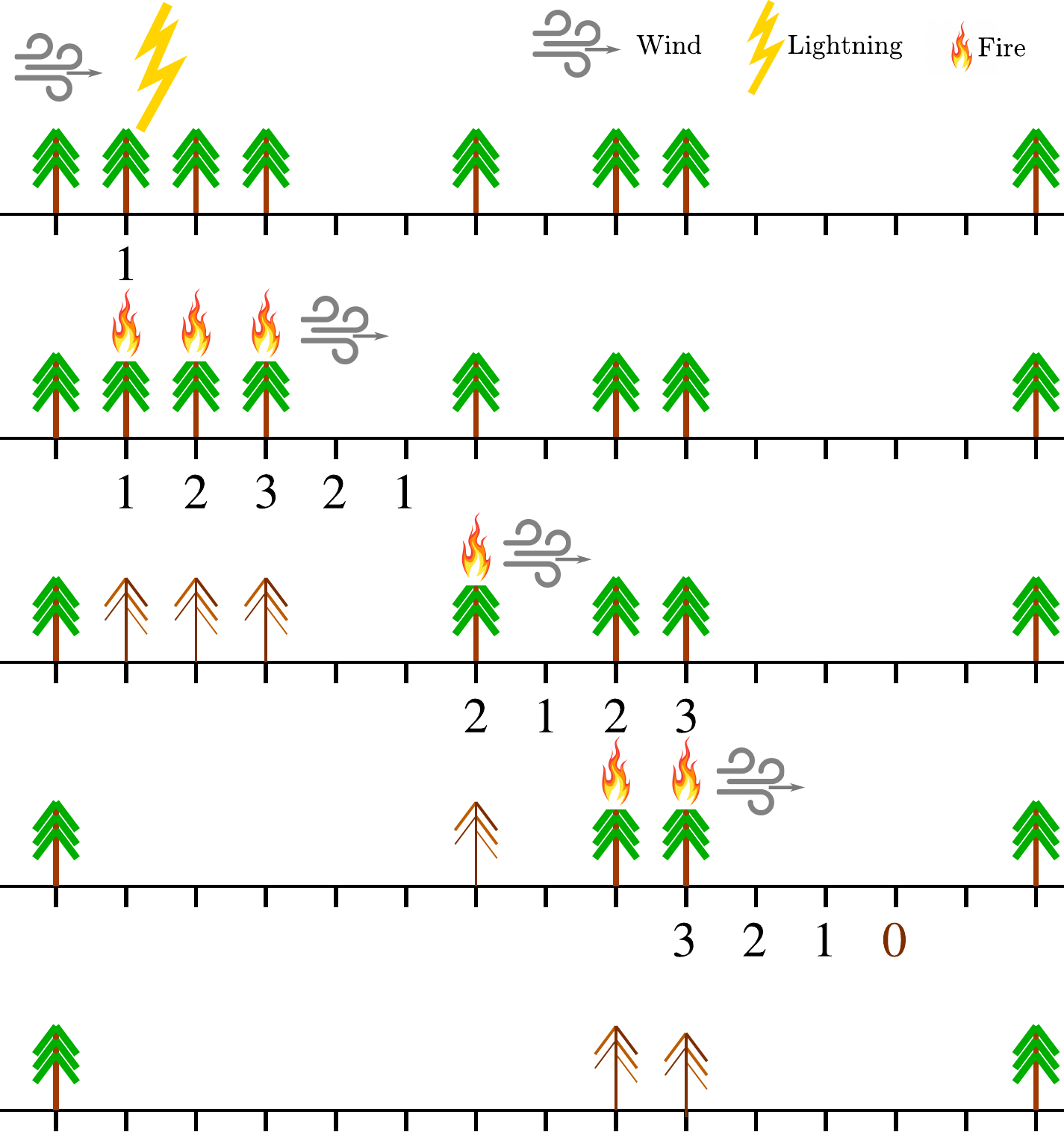}
\caption{Wind-driven forest fire model. Each row shows the system state after each forest has burned.  After the lightning strike on the second tree of a forest of size 4, a fire of intensity 3 is created. The numbers indicate the fire intensity at each site. The fire jumps the first gap because its intensity exceeds the gap length. The fire stops when its intensity reaches zero. }
\label{fig:model}
\end{figure}

\paragraph*{\textbf{Model:}} We treat a periodic one-dimensional ring of length $L$. Each site is either empty or occupied by a tree. A constant wind helps spread fire to downwind trees and to jump gaps between forests (Fig.~\ref{fig:model}). If the initial forest has $I$ trees downwind from the lightning strike, the fire has intensity $I$ at the downwind edge of this forest. If the adjacent downwind gap length is $G$ and $I \leq G$, the fire stops. If $I>G$, the fire jumps the gap to reach the next forest with diminished intensity $I-G$ because the fire intensity decreases by 1 for each empty site jumped. As the fire burns through this next forest, its intensity increases by 1 for each tree burned. At the downwind edge of this forest, the same gap-jumping rule is again applied. The fire stops when its intensity reaches zero. All steps in the fire propagation are regarded as instantaneous.

\begin{figure}[ht]
    \centering
\subfigure[$\phantom{+}f=0.0001$]{\includegraphics[width=\linewidth]{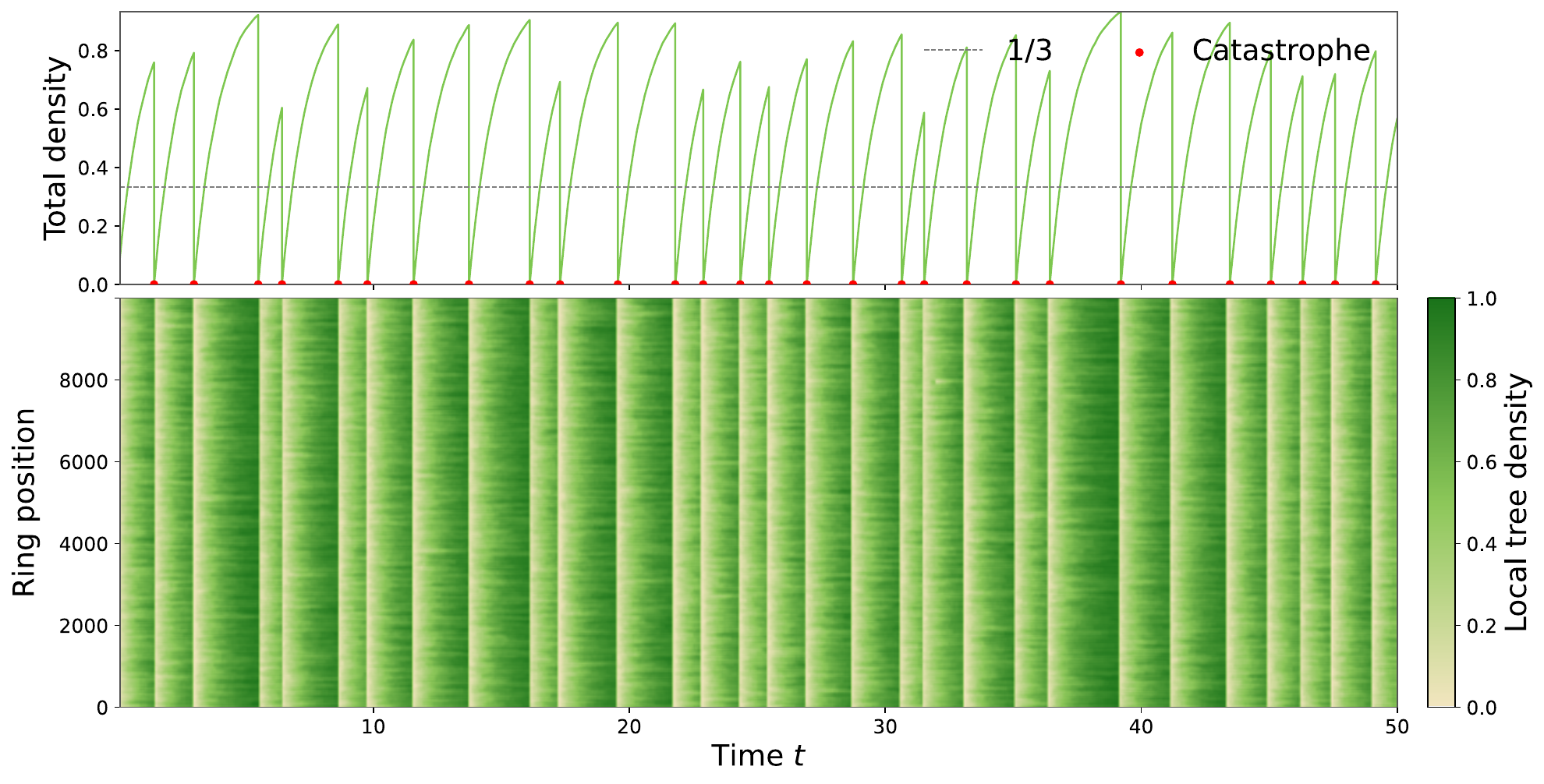}}
\subfigure[$\phantom{+}f=0.2$]{ \includegraphics[width=\linewidth]{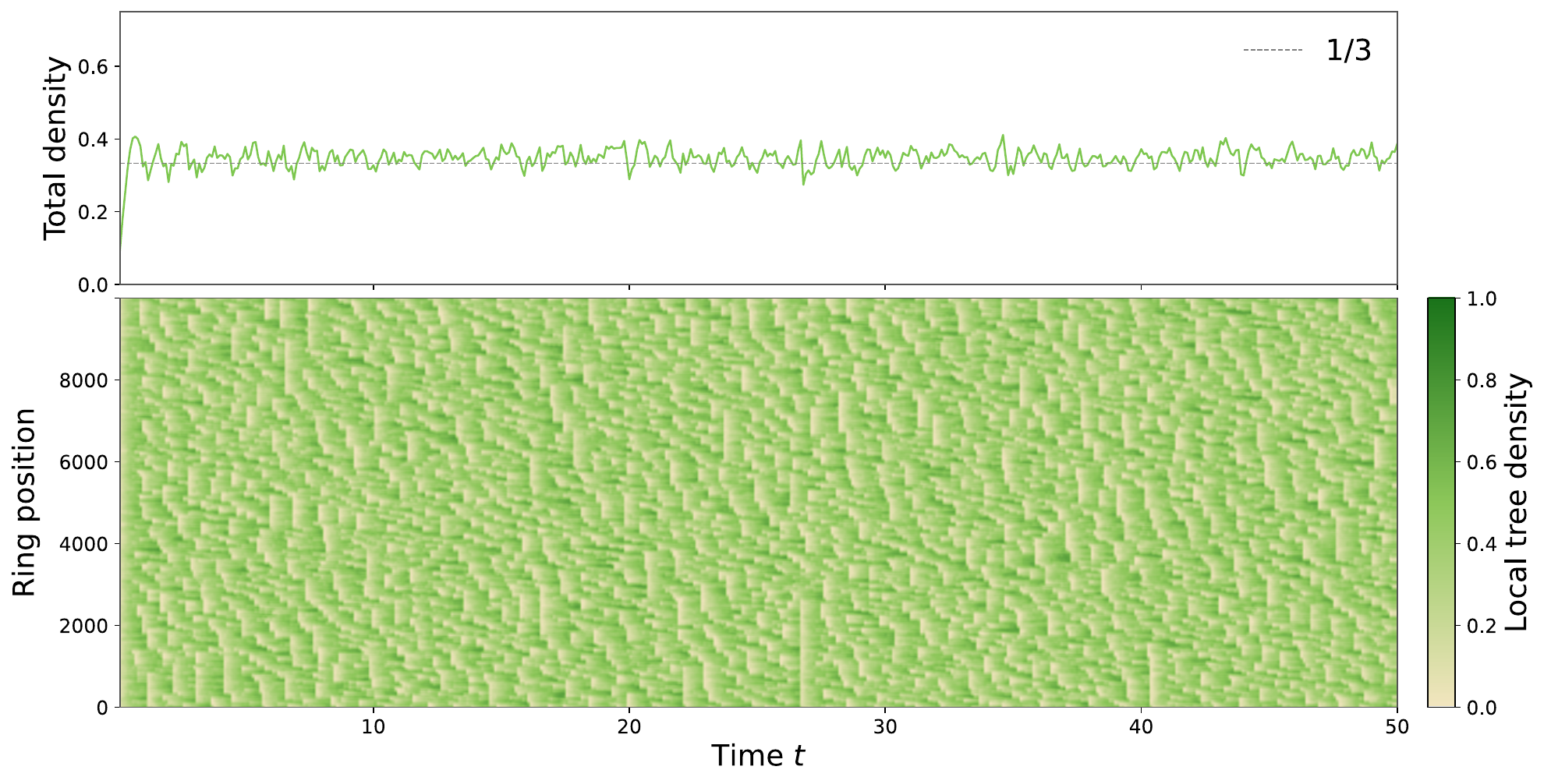}}
\subfigure[$\phantom{+}f=0.01$]{ \includegraphics[width=\linewidth]{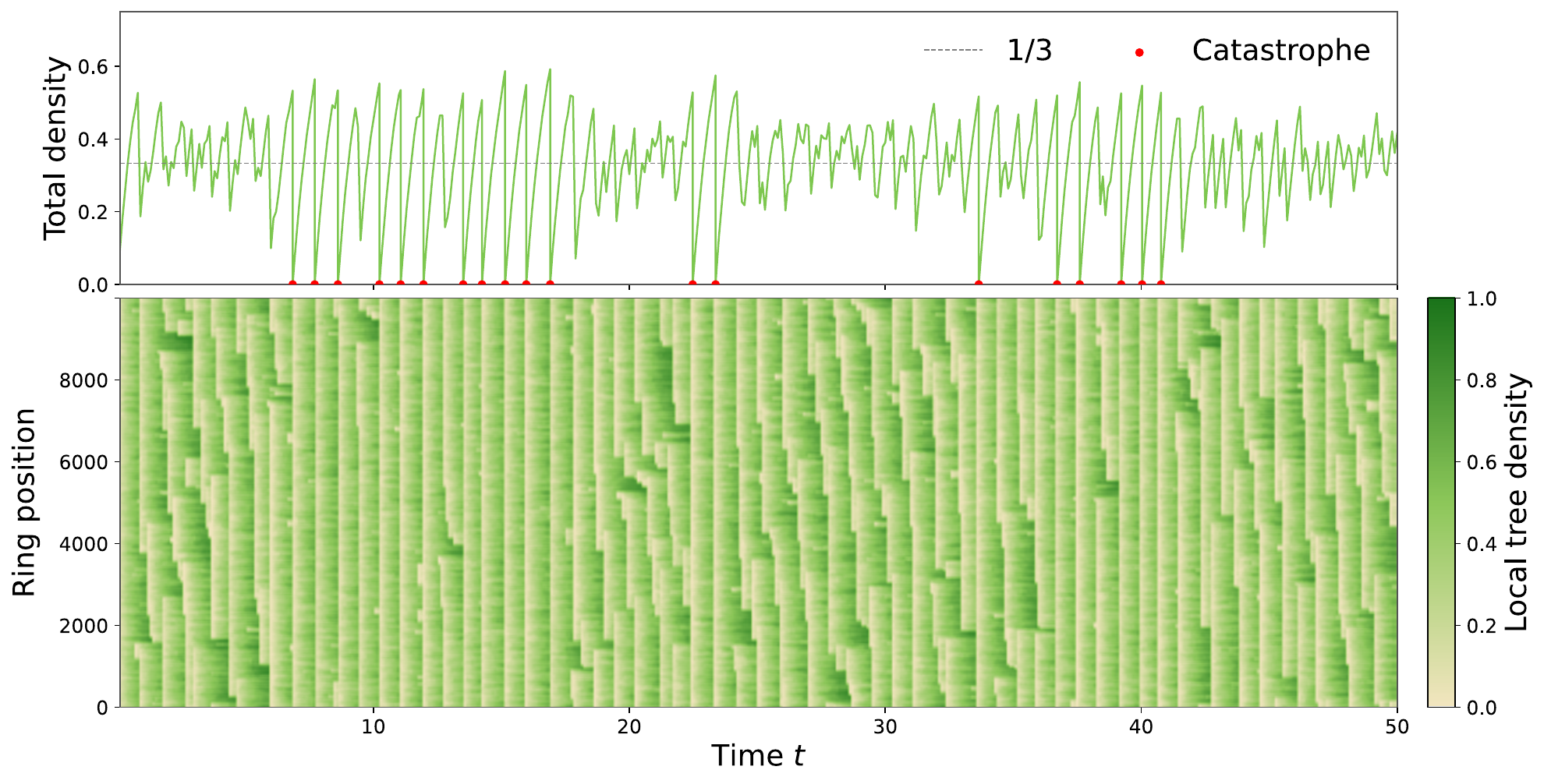}}
    \caption{Spacetime plots of tree dynamics on a ring of length $L=10^4$.  (a) Lightning rate $f=10^{-4}$, which corresponds to the quasi-deterministic regime. (b) $f=2\times 10^{-1}$, which corresponds to the steady state.  (c) $f=10^{-2}$, which corresponds to the intermittent regime. The upper curve in each panel shows the time dependence of the tree density. The bottom plots show the local tree density over space estimated with a running average over windows of length 100 at fixed time intervals.}
    \label{fig:spacetime}
\end{figure}

Our choice of a unit intensity increase/decrease as the fire propagates in a forest or across a gap is motivated by simplicity and its natural connection to first-passage in one dimension~\cite{Redner2001FirstPassage,BMS13}. By viewing the intensities in Fig.~\ref{fig:model} as the position of an equivalent 1d random walk, extinguishing a fire corresponds to the random walk first reaching the origin.  This connection suggests that a particularly interesting case should be equal intensity gains/decreases as the fire propagates through forests/gaps and tree density close to 1/2; this case corresponds to an unbiased random walk.

We emphasize that our wind-driven reinforcement mechanism is starkly different than earlier forest-fire models that attempted to account for the effect of wind by introducing anisotropy in the fire propagation mechanism~\cite{OhtsukiKeyes1986,vonNiessenBlumen1986,vonNiessenBlumen1988,DuarteCarvalhoRuskin1992,Drossel1994ZNatA}.  The reinforcement mechanism in our model is also distinct from the mechanisms in various long-range spreading phenomena, such as in seed dispersal~\cite{Clark1999,Levin2003,Nathan2006}, transmission of pathogens and diseases~\cite{BrownHovmoller2002,McCallum2003,Grassberger2013JSME,Grassberger2013JSP,BrockmannHelbing2013}, human mobility~\cite{Brockmann2006,Gonzalez2008,Rhee2011}, and the spread of mutations~\cite{HallatschekFisher2014}.  In all these examples, the nature of the long-range transport is independent of the extent of the dispersal.  In contrast, the fire intensity in our model either increases or decreases depending on the forest and gap geometry.  In this sense, our model resembles that of Ref.~\cite{HebertDufresne2025PRL}, where the intensity of a spreading attribute also increases or decreases by a self-reinforcement mechanism.

\paragraph*{\textbf{The three regimes:}} Our model exhibits rich dynamics, with three distinct regimes as a function of the lightning rate $f$: (a) quasi-deterministic, (b) steady state, and (c) intermittent (Fig.~\ref{fig:spacetime}). In the quasi-deterministic regime, lightning is sufficiently rare that the tree density reaches a value greater than 1/2 before lightning strikes. When lightning finally does strike, the subsequent fire consumes all trees because gaps are typically smaller than forests. We call fires that span the entire system \textit{catastrophes}. The outcome is the sawtooth-like time dependence of the tree density shown in Fig.~\ref{fig:spacetime}(a). 

We now estimate the range of $f$ where quasi-determinism arises. Between successive fires, the tree density $\rho(t)$ grows according to $\dot\rho=1\!-\!\rho$, with solution $\rho(t)=1\!-\!e^{-t}$ for the initial condition $\rho(0)\!=\!0$. Once the tree density $\rho(t)$ reaches 1/2, which happens at  $t_{1/2}=\ln 2$, the ensuing fire is almost surely a catastrophe. Starting from an empty system at $t=0$, the number of trees at time $t$ equals $\rho(t)L$, so that the rate of lightning strikes on occupied sites is $f\rho(t)L$.  The integrated hazard up to time $t_{1/2}$ thus is
\begin{align}
  \Lambda = \int_0^{t_{1/2}}f\rho(t)\,L\, dt = fL(\ln 2 -\tfrac{1}{2})\,.
\end{align}
Since lightning strikes occur by a Poisson process, and approximating the much faster increase in the tree density as continuous, the probability $\Pi$ that no lightning strikes occur before the tree density reaches 1/2 is $\Pi\approx e^{-\Lambda}$. We define the end of the deterministic regime by the criterion $\Pi=1/2$.  Using the above expression for $\Lambda$, we find
\begin{align}
  f_{\rm det} = \frac{\ln 2}{\ln 2-\frac{1}{2}}\frac{1}{L}\approx \frac{3.589}{L}\;.
\end{align}
For lightning rates $f<f_{\rm det}$, the tree density is greater than 1/2 when lightning strikes, so that all trees are consumed in the ensuing fire. This dynamics leads to the time dependence of the tree density shown in Fig.~\ref{fig:spacetime}(a).

\begin{figure}
    \centering
    \includegraphics[width=0.9\linewidth]{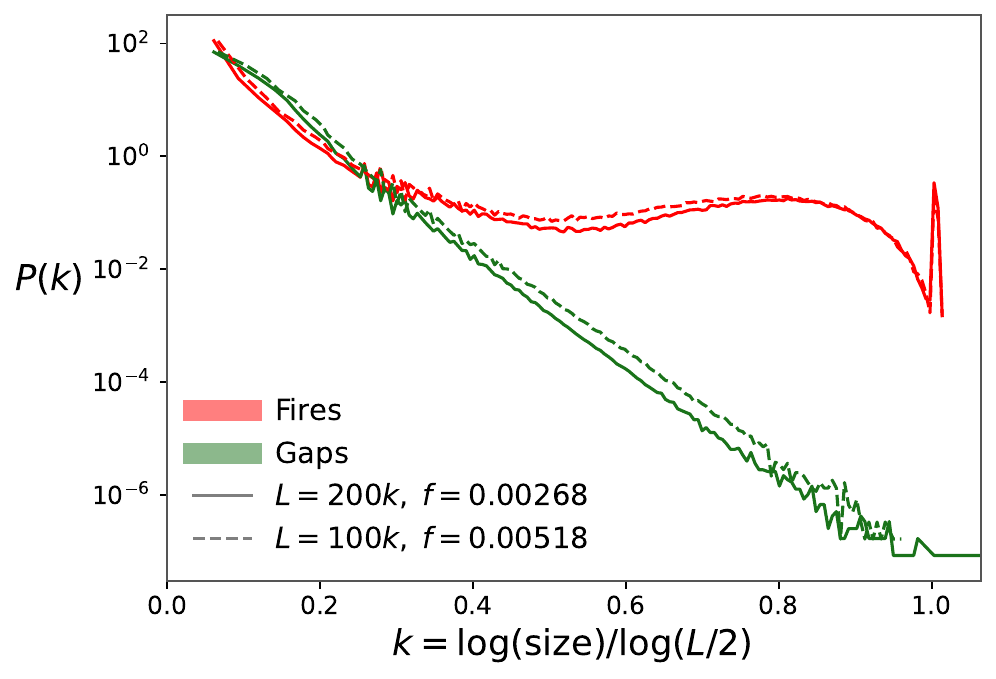}
    \caption{The distribution of the log of the fire sizes (red) and gap sizes (green) for systems of different sizes $L$ and different lightning rates $f$. The data are scaled so that $k=1$ corresponds to a size of $L/2$ and then smoothed by logarithmic binning. The gap size distribution in the intermittent regime is a robust power-law distribution with scaling exponent close to $-3$. }
    \label{fig:dist}
\end{figure}

The statistics and dynamics of forest fires are particularly intriguing in the intermittent regime (Fig.~\ref{fig:spacetime}(c)). Figure~\ref{fig:dist} shows the distribution of fire and gap sizes for systems of different sizes with $f$ chosen to be within the intermittent regime of the fire dynamics. The data have been smoothed by logarithmic binning and the abscissa has been rescaled so that $k=1$ corresponds to a size $L/2$; this is generally the minimum size for a fire to be a catastrophe. There are three noteworthy aspects of the fire size distributions. First, there are sharp peaks in the distributions at the large-size extreme; these correspond to catastrophes. For small sizes there is a monotonic decay in the distribution that is roughly consistent with power-law behavior. Unexpectedly, there are intermediate peaks that arise from fires within the chaotic temporal windows of Fig.~\ref{fig:spacetime}(c) for scaled abscissa $k^*\approx 0.8$.  These peaks correspond to typical fire sizes that scale as $L^{k^*}$. The exponent of $-3$ for the gap-size distribution agrees with previous results for the forest-fire model without wind~\cite{Drossel1993exact}.

\begin{figure}
    \centering
    \includegraphics[width=\linewidth]{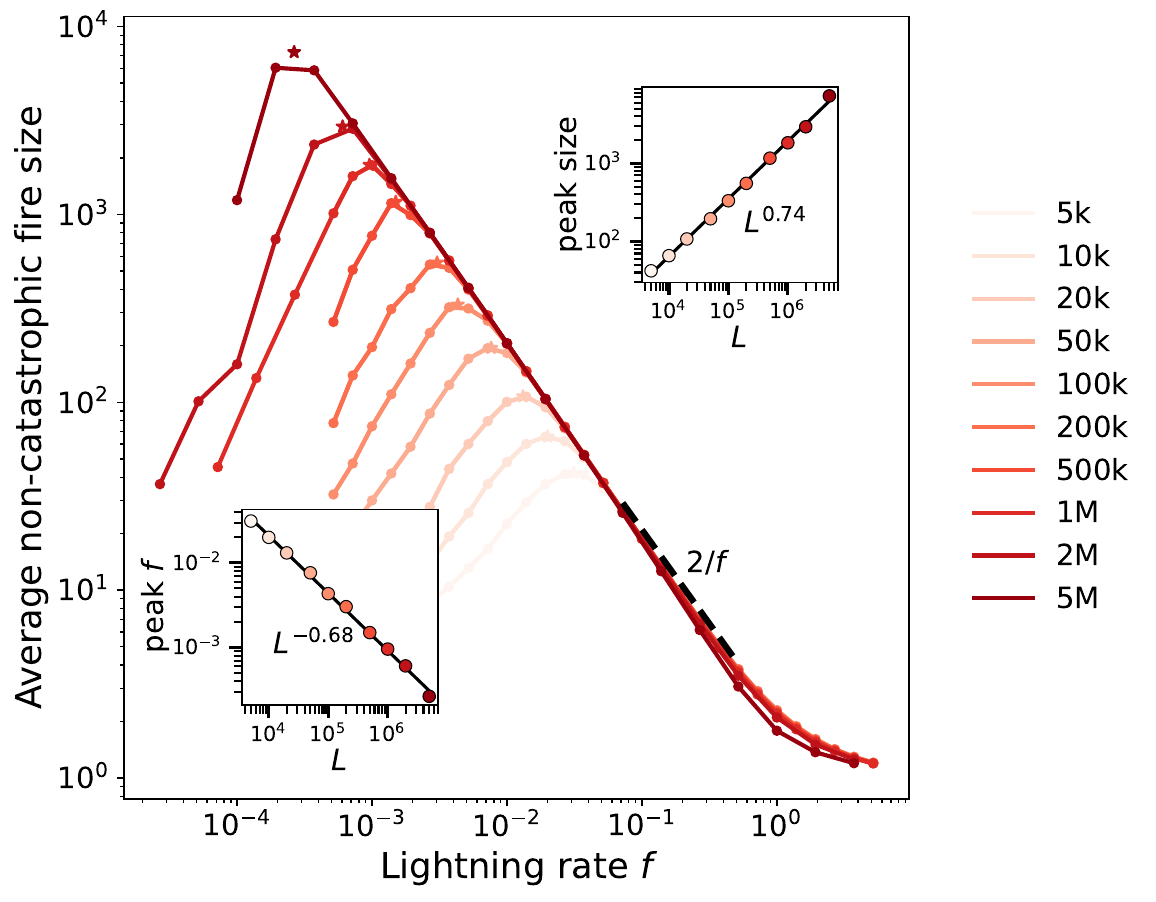}
    \caption{Characteristic fire size, excluding catastrophes. The stars indicate the peak fire size as inferred from a parabolic fit to the curves near the maximum. Insets show the peak fire size (top right, fit with $0.07L^{+0.74}$) and the corresponding value of $f$ (bottom left, fit with $10.9L^{-0.68}$) as a function of system size $L$. Both fits have error bars of 0.02.}
    \label{fig:average_fire}
\end{figure}

To better quantify these intermediate-size fires, Fig.~\ref{fig:average_fire} shows the dependence of the average fire size on $f$ when catastrophes are excluded from the average. Starting at a large value of $f$ that corresponds to the steady state regime, the mean fire size increases as $f$ is decreased until catastrophes arise.  Subsequently, the average size of fires decreases because of the exclusion of the large contribution due to catastrophes from the average. We find that the peak fire size scales as $L^{0.74\pm 0.02}$ (upper inset to Fig.~\ref{fig:average_fire}) and this peak is located at a value of $f$ that scales as $L^{-0.68\pm 0.02}$ (lower inset to Fig.~\ref{fig:average_fire}). While the fire size distribution is characterized by multiple scales in the intermittent regime, the gap sizes follow a robust power-law distribution with scaling exponent of $-3$. This estimate is based on the methods of Ref.~\cite{HebertDufresne2025PRL} and arises for all $f$ values studied between $10^{-5}$ and $10^{-1}$. It is striking that self-organized criticality arises in the gap sizes but not fire sizes.

From the simulation data in Fig.~\ref{fig:average_fire}, we now estimate the value of $f$ that separates the quasi-deterministic and steady-state regimes. We expect that intermittency arises when catastrophes occur only infrequently. This condition requires that the expected total tree growth rate matches the expected total rate at which trees burn when tree density $\rho$ is close to $1/2$; this is the minimal tree density that allows catastrophes to occur. At $\rho=1/2$, the total growth rate is $L/2$. The rate at which fires are ignited is $fL/2$ and at the critical point, their expected size is roughly $0.07\times L^{0.74}$ (right inset of Fig.~\ref{fig:average_fire}). Equating these growth and burn rates gives the following estimate for the lightning rate that leads to intermittency:
\begin{align}
  f_{\rm int}\sim 14.3/ L^{0.74}\;,
\end{align}
which roughly matches the $f$ values in the left inset of Fig.~\ref{fig:average_fire} and serves as a useful consistency check.

\begin{figure}[t!]
\centering
\includegraphics[width=0.8\linewidth]{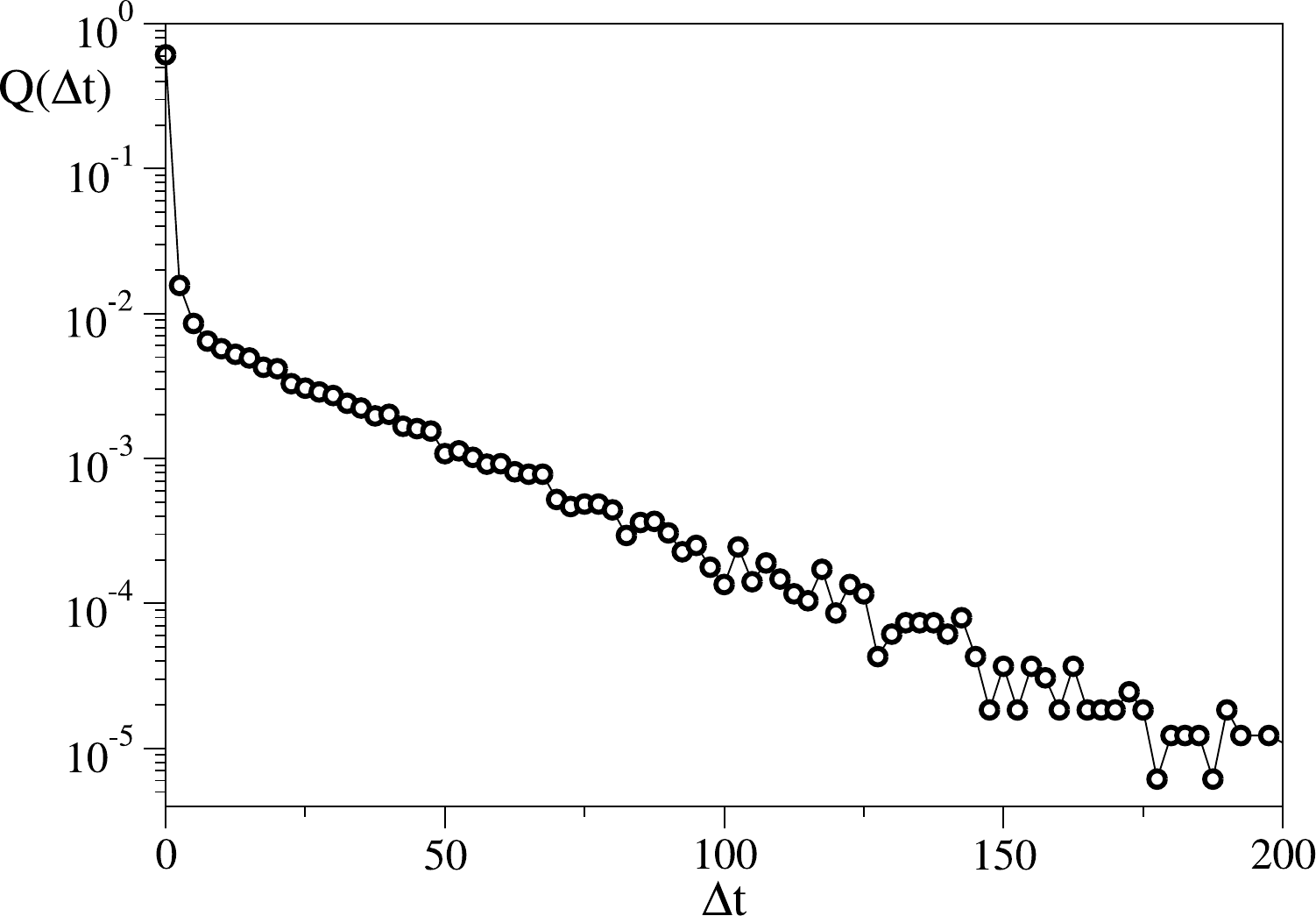}
\caption{The probability distribution $Q(\Delta t)$ of the time intervals $\Delta t$ between catastrophes for the case of $L=10^5$ and $f=3\times 10^{-3}$.}
\label{fig:dt}
\end{figure}

\begin{figure*}[ht]
\centering
\subfigure[]{\includegraphics[width=0.325\linewidth]{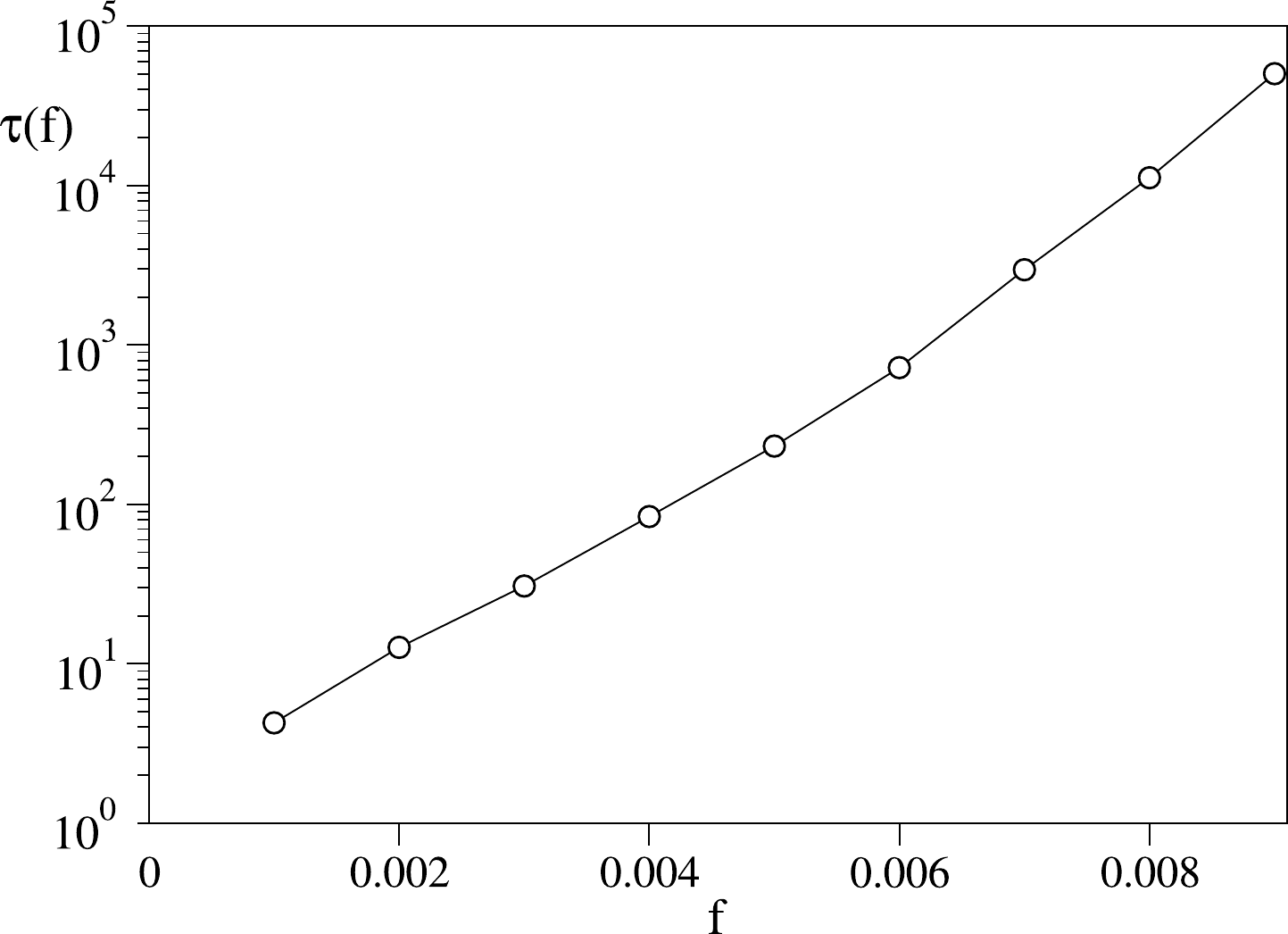}}
\subfigure[]{\includegraphics[width=0.325\linewidth]{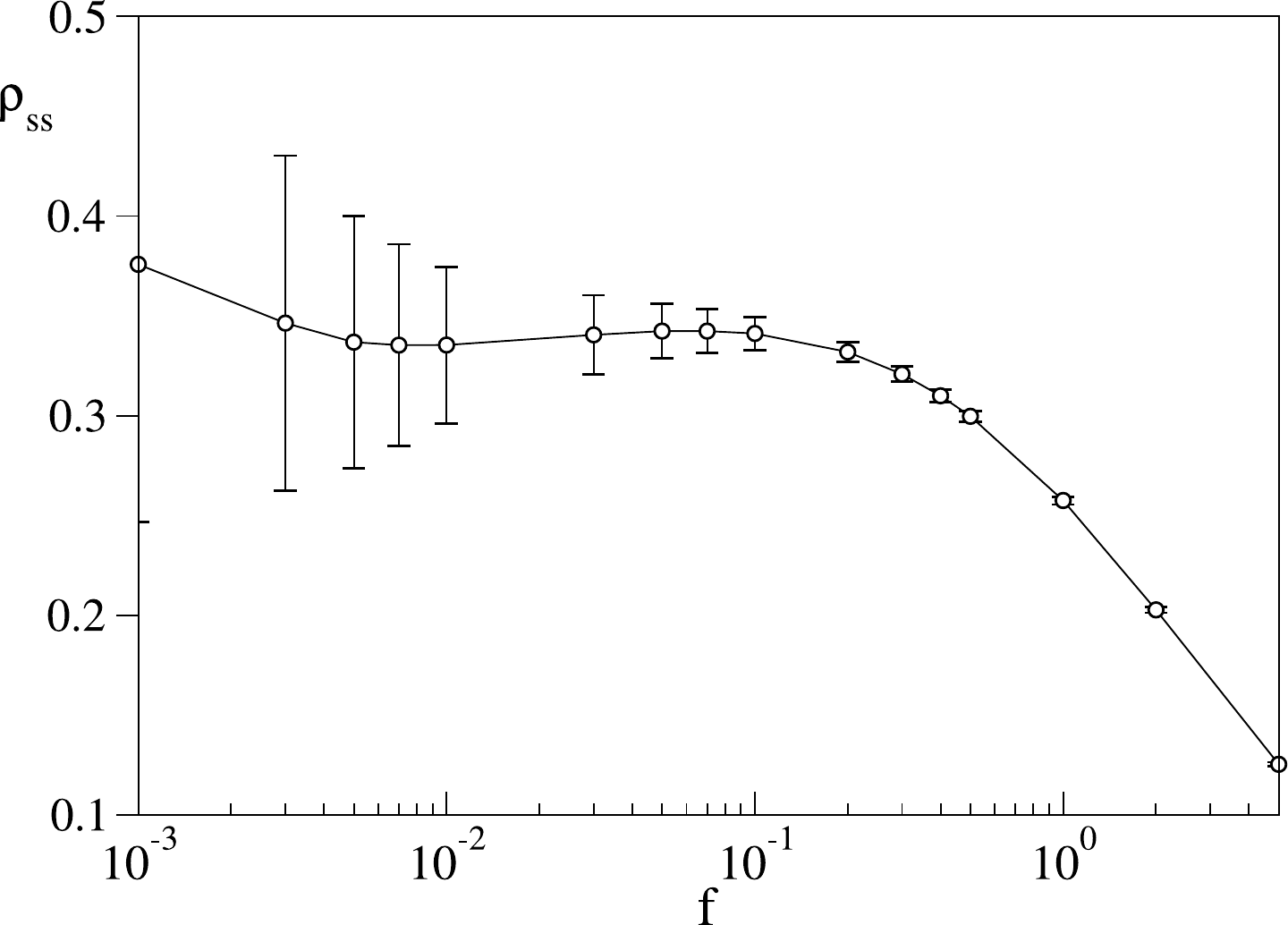}}
\subfigure[]{\includegraphics[width=0.325\linewidth]{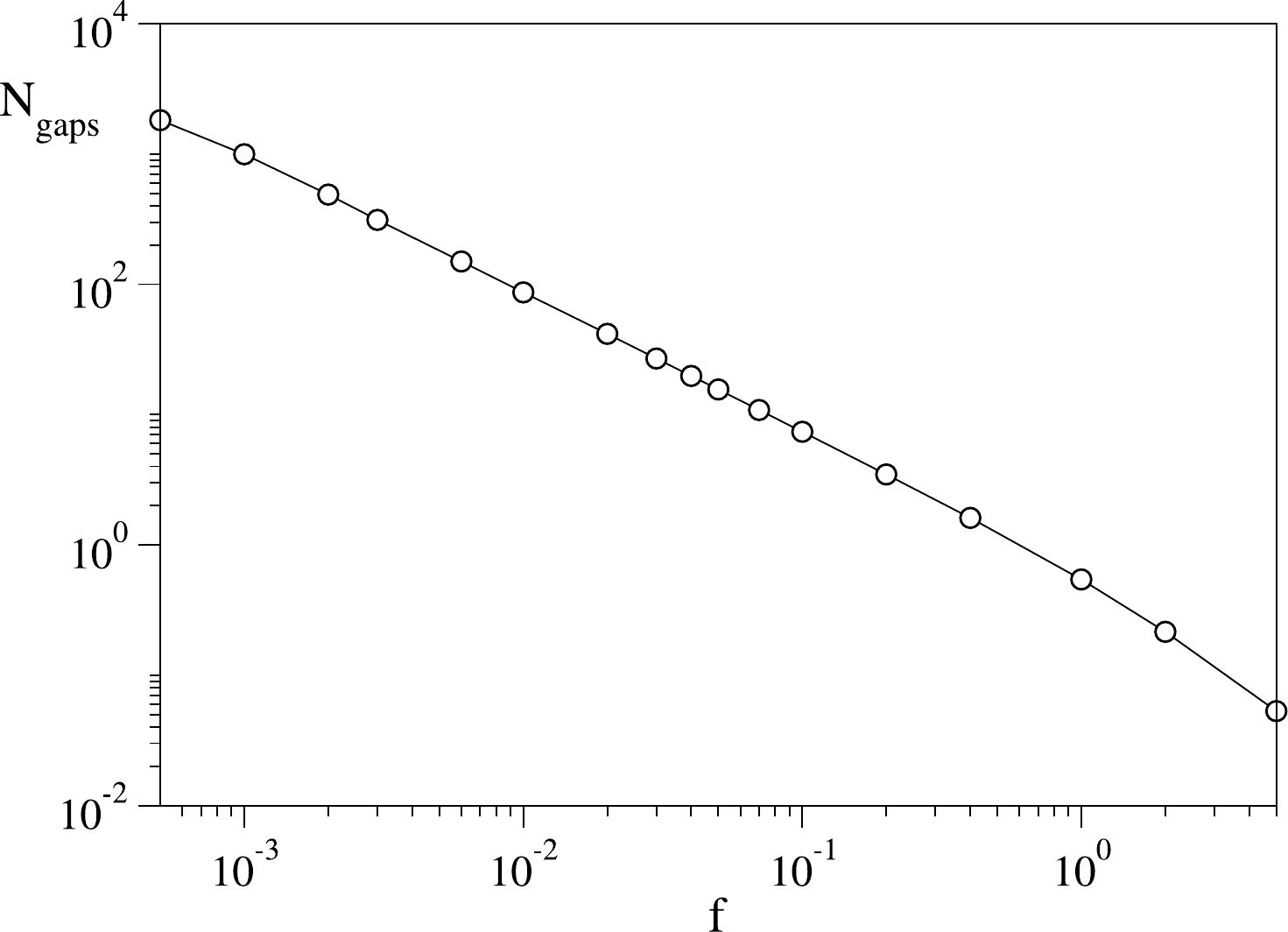}}
\caption{(a) The time scale $\tau(f)$ associated with the exponential decay of $Q(\Delta t)$ in Fig.~\ref{fig:dt}. (b) Steady-state tree density versus $f$. (c) Average number of gaps jumped by fires versus lightning rate $f$. All data are based on a system of length $L=10^5$.}
\label{fig:ss}
\end{figure*}

A fundamental characteristic of the intermittent dynamics is the distribution $Q(\Delta t)$ of time intervals $\Delta t$ between catastrophes. We expect $Q(\Delta t)$ to be sharply peaked in the deterministic temporal windows, such as between $t=13$--17 in Fig.~\ref{fig:spacetime}(c). We also expect it to reflect a multiplicity of time scales when chaos occurs, such as $t$ between 18--22, 22--33, and 40 onward in Fig.~\ref{fig:spacetime}(c). Figure~\ref{fig:dt} illustrates that this multiscaling is characterized by two time scales: (a) an interval of the order of 1 that corresponds to the time between catastrophes in the nearly deterministic temporal windows and (b) a scale $\tau(f)$ that is associated with time intervals between catastrophes when the intervening dynamics is chaotic. We extract this second scale from the clearly defined exponential decay of $Q(\Delta t)$ in Fig.~\ref{fig:dt}. Similar two time-scale behavior to that shown in this figure occurs for all other values of $f$ in the intermittent regime.

The dependence of $\tau(f)$ is rather surprising. As $f$ is increased from the intermittent towards the steady-state regime, catastrophes become progressively rarer. We might therefore expect that a point would be reached where catastrophes no longer occur. The data in Fig.~\ref{fig:ss}(a) suggests otherwise. Here the dependence of $\tau(f)$ appears to be either exponential or slightly faster than exponential. Catastrophes can occur for any value of $f$, but the waiting time between catastrophes becomes astronomically long as $f$ increases towards the steady-state regime.

We now turn to the dynamics of the steady state of Fig.~\ref{fig:spacetime}(b). A striking feature of this state, shown in Fig.~\ref{fig:ss}(b), is that the tree density is nearly constant over a wide range of $f$ values.  We estimate the steady-state density from the scaling relationship between average fire size and $f$ observed in Fig.~\ref{fig:average_fire}. Namely, for $f>f_{\rm int}$, the characteristic fire size is independent of $L$ and scales as $2/f$. We now determine the steady-state $\rho_{\rm ss}$ by equating the total burn rate with the tree growth rate. Fires are ignited at rate $f\rho(t)L$ and burn $2/f$ trees. This gives a total tree burning rate equal to $2\rho(t)L$. The total tree growth rate is simply $\left[1-\rho(t)\right]L$.   Equating these two rates, we determine the steady-state tree density $\rho_{\rm ss}$ to be
\begin{equation}
    \rho_{\rm ss} = \frac{1}{3}\,.
\end{equation}  
This simple prediction is quite close to our simulation results for a system of size $L=10^5$ and for $f$ in the range $3\times 10^{-3}\leq f\leq 3\times 10^{-1}$ (Fig.~\ref{fig:ss}(b)).  When $f$ is decreased below $10^{-3}$, density fluctuations become of the order of the density itself, so that its average no longer meaningfully quantifies forest statistics.  This is also the regime where most of the fires are catastrophes so that the system is no longer in the steady state.

The steady-state regime further divides into two subregimes in which fires are either: (i) local or (ii) long ranged. We define a fire as local when the average number of gaps that a fire jumps is less than 1. In this subregime, forests are necessarily small and surrounded by large empty spaces, so that fires typically cannot propagate beyond the initial forest. Wind therefore plays almost no role in the fire propagation. Numerically, we find that primarily local fires occur when $f\agt 1/2$ (Fig.~\ref{fig:ss}(c)).

On the other hand, fires can spread nonlocally when $f_{\rm int}<f<1/2$. In spite of this nonlocality, the dynamics remains steady. For this range of $f$, we numerically find that the average number of gaps $N_{\rm gaps}$ jumped by a fire is well fit by the form $N_{\rm gaps}\sim A f^{-\nu}$, with $A\approx 0.8$ and $\nu\approx 1.03$ (Fig.~\ref{fig:ss}(c)). In spite of this increasing nonlocality of the fire propagation, the average tree density still varies within a narrow range, as shown in Fig.~\ref{fig:ss}(b).

\paragraph*{\textbf{Discussion:}} We introduced a simple one-dimensional forest-fire model in the spirit of self-organized criticality. The new feature of our model is the presence of a steady wind that allows the fire to jump gaps between disjoint forests by a self-reinforcing mechanism.  Despite this new mechanism, the system maintains a signature of self-organized criticality in the form of a power-law distribution of gap sizes (not fire sizes) with a robust scaling exponent of $-3$.  Self-reinforcement leads to rich phenomenology with three disparate regimes of temporal behavior. Quasi-determinism arises when the lightning rate $f$ is of the order of $1/L$ or smaller. Here, all fires are catastrophes and the tree density has a sawtooth-like time dependence. There is a steady state when lightning strikes are sufficiently frequent that fires are small even when a fire jumps multiple gaps.  In this steady state, the tree density is close to 1/3 over a wide range of $f$ values, even when fires are not local.  

Most strikingly, there exists a range of lightning rates $f$, with $f\sim L^{-\mu}$ and $\mu\approx 0.8$,  where intermittent behavior arises. This intermittency is characterized by temporal domains of nearly deterministic cycles of steady tree growth followed by catastrophes that are interspersed with chaotic temporal domains with no discernible systematic patterns of behavior. In this intermittent regime the distribution of fire sizes is characterized by multiple scales.  There are catastrophes, whose size scales linearly with $L$, as well as intermediate size fires within the chaotic windows whose size scales as $L^{\mu}$. These macroscopic but non-catastrophic fires create firebreaks that hinder the re-establishment of catastrophes.

While the phenomenology of the model is quite rich, our theoretical understanding is meager. It will be worthwhile to develop analytical approaches that will complement the simulation results reported here.  It will be interesting to explore whether the phenomena of our model will also occur for different rules for the growth and diminishment of the fire intensity. Another important and more practical direction is to extend the model to two dimensions and attempt to make connections with real forest fires in the presence of wind or other sources of non-locality, such as terrain heterogeneities, an aspect of forest fires that has very recently been investigated~\cite{Florez2026Topographic}.

\paragraph*{\textbf{Acknowledgments}.}
The authors acknowledge helpful discussions with P. L. Krapivsky. This collaboration was supported by the Santa Fe Institute.

\bibliographystyle{apsrev4-2}
\bibliography{fire.bib}
\end{document}